\documentclass{svproc}
\usepackage{url}

\usepackage{graphicx}
\usepackage{booktabs}
\usepackage{threeparttable}
\usepackage{multirow}
\usepackage{multicol}
\usepackage{pifont}
\usepackage{subcaption}
\usepackage{color}
\usepackage{framed}
\usepackage{booktabs}
\usepackage{graphicx}
\usepackage{amsmath,amssymb,amsfonts}
\usepackage{dirtytalk}
\usepackage{hyperref}
\usepackage{comment}
\usepackage{enumerate}
\usepackage{algorithm,algpseudocode}
\usepackage{listings}

\newcounter{t0d0_counter}
\newcommand{\nofixme}[1]{
}

\newcommand{\specialcell}[2][c]{%
  \begin{tabular}[#1]{@{}c@{}}#2\end{tabular}}

\newcommand{\cctvexposure}{\texttt{CCTV-Exposure}}

\begin{document}
\mainmatter              
%
\title{\cctvexposure{}: An open-source system for measuring user's privacy exposure to mapped CCTV cameras based on geo-location (Extended Version)\thanks{Authors' extended version of peer-reviewed paper first published in ``Business Modeling and Software Design: 12th International Symposium, BMSD 2022, Fribourg, Switzerland, June 27-29, 2022, Proceedings'' by Boris Shishkov (Editor) by Springer Nature''~\cite{turtiainen2022bmsd}.} }
\titlerunning{\cctvexposure{}}  
%
\author{Hannu Turtiainen\inst{1}, Andrei Costin\inst{1}\thanks{Corresponding and original idea author.}, \and Timo H\"{a}m\"{a}l\"{a}inen\inst{1}}
\authorrunning{H. Turtiainen, A. Costin, T. H\"{a}m\"{a}l\"{a}inen} 
%
\tocauthor{Hannu Turtiainen, Andrei Costin, and Timo H\"{a}m\"{a}l\"{a}inen}
\index{Turtiainen, H.}
\index{Costin, A.}
\index{H\"{a}m\"{a}l\"{a}inen, T.}

\institute{Faculty of Information Technology, University of Jyv\"{a}skyl\"{a}, P.O. Box 35, Jyv\"{a}skyl\"{a}, 40014 Finland
\email{\{turthzu,ancostin,timoh\}@jyu.fi}
\url{https://jyu.fi/it/}
}

\maketitle              

\begin{abstract}

In this work, we present \cctvexposure{} -- the first CCTV-aware solution to evaluate potential privacy exposure to closed-circuit television (CCTV) cameras. 
The objective was to develop a toolset for quantifying human exposure to CCTV cameras from a privacy perspective. 
Our novel approach is trajectory analysis of the individuals, coupled with a database of geo-location mapped CCTV cameras annotated with minimal yet sufficient meta-information. 
For this purpose, \cctvexposure{} model based on a Global Positioning System (GPS) tracking was applied to estimate individual privacy exposure in different scenarios.
The current investigation provides an application example and validation of the modeling approach. 
The methodology and toolset developed and implemented in this work provide time-sequence and location-sequence of the exposure events, thus making possible association of the exposure with the individual activities and cameras, and delivers main statistics on individual's exposure to CCTV cameras with high spatio-temporal resolution.


\end{abstract}
\section{Introduction}
\label{sec:intro}

In the modern world, public spaces of many cities are being surveilled by closed-circuit television (CCTV) cameras to a considerable extent. It is estimated that globally there are around 1 billion CCTV cameras in use today~\cite{bischoff-2021,cnbc2019billion}. As an example, there are approximately half a million CCTV cameras in London, and an average person living there is recorded on camera 300 times everyday~\cite{caughtoncamera2019cctvlond}. In the United States, people are likely recorded by a CCTV camera over fifty times per day~\cite{ipvm2016us}. In 2019, a person documented 49 CCTV cameras on the way to work in New York City~\cite{bi2019cctv} and described it as dystopian. 

The discourse on CCTV surveillance has ethical dimensions. Von Hirsch argues that CCTV surveillance is sometimes covert, and often people believe that they are not under CCTV surveillance when they are~\cite{vonhirsch2000ethics}. Furthermore, according to a 2016 survey, an average citizen of the United States is assumed to be recorded by four or fewer CCTV cameras per day, while the actual figure is likely over ten times larger~\cite{ipvm2016us}. Considering the amount of CCTV cameras having been installed globally and the fact that people can be detected and recorded by them, adding face recognition to the pattern opens up an unsettling possibility to also automatically identify people by CCTV cameras~\cite{hu2004survey,wheeler2010face,axis-id-rec}. Overall, there is controversy about the ethics of CCTV surveillance, and books have been written about the subject~\cite{vonhirsch2000ethics,larsen2011setting}. Moreover, CCTV cameras, Digital Video Recorders (DVRs), and Video Surveillance Systems (VSSs) are notoriously known to be vulnerable to cybersecurity attacks and hacks~\cite{costin2016security}. Therefore, it is reasonable to assume that the CCTV cameras overlooking public places may be under the control of unauthorized persons hence posing a direct threat to privacy. 

In this context, we argue that it is essential to create \emph{CCTV-aware} solutions and technologies that allow people the discretion to be under surveillance or not in public places.
We approach the question from the perspective of estimating individual users' exposure to CCTV cameras based on their real-time or historical geo-location (e.g., position, tracks, routes). 
While there is a substantial amount of studies related to exposure to various ``harmful environments''~\cite{dias2014modelling,tchepel2014modeling,beekhuizen2013performance,breen2014gps,ma2020assessing,hinton2019gps}, to the best of our knowledge, none of the existing works focuses on the exposure to privacy invasion by CCTV cameras when this is seen as a ``harmful environment'' for individual privacy. 

Furthermore, as positioning systems such as the GPS are a ubiquitous technology in today's modern world, where smartphones and sporting gadgets have GPS built-in and many software track users' movements, from a privacy standpoint, one can argue that the users mishandle their GPS and location data and compromise their privacy. 

Nevertheless, when shared responsibly and for practical purposes, the users' GPS data can also be used for Privacy Enhancing Technologies (PET), as we present in this paper. 
In this paper, we propose one such \emph{CCTV-aware} solution, namely \cctvexposure{}. 
When compared to exposure to ``harmful environments'' such as exposure to radiation, the \cctvexposure{} system is intended to act like a ``CCTV dosage meter'' for travel activities of privacy-minded individuals. 

Our contributions with this work are:
\begin{enumerate}
\item We are the first (to the best of our knowledge) to propose, implement, and demonstrate a system aimed at measuring individuals' privacy exposure to CCTV cameras using analysis of historical and real-time GPS data
\item For evaluation and further improvements, we release (upon peer-review acceptance) the relevant artifacts (e.g., code, data, documentation) as open-source: \url{https://github.com/Fuziih/cctv-exposure}
\end{enumerate}

The rest of this paper is organized as follows. 
We briefly introduce related work in Section~\ref{sec:relatedwork}. 
We present in Section~\ref{sec:impl} our algorithms as well as design and implementation details. 
Then, in Section~\ref{sec:results} we introduce results and their evaluation. 
Finally, we conclude this paper with Section~\ref{sec:concl}. 

\section{Related Work}
\label{sec:relatedwork}

To date, to the best of our knowledge, none of the works (systems, implementations, surveys) have 
addressed the research question related to individuals' privacy exposure to CCTV and video surveillance, as we do in this paper. 
However, we briefly introduce closely related state-of-the-art and related work in adjacent fields below. 

Turtiainen et al.~\cite{turtiainen2020towards} were the first to propose and develop a dedicated computer vision (CV) model -- CCTVCV -- designed specifically 
to detect CCTV cameras from street view and other images, with the primary intended purpose of building various 
privacy-enhancing technologies (PET), tools, and large-scale datasets (e.g., global mapping of CCTV cameras in public spaces). 
Building on the applicative ideas from Turtiainen et al.~\cite{turtiainen2020towards}, 
Sintonen et al.~\cite{sintonen2021osrm} developed and proposed OSRM-CCTV, which is the first of its kind route planning and management. 
PET solution offers pro-active route planning optimized for individual privacy and public safety. 
Subsequently, Lahtinen et al. ~\cite{lahtinen2021towards} applied and validated an early prototype of OSRM-CCTV 
to demonstrate the feasibility of OSRM-CCTV in real cities (e.g., Jyv\"{a}skyl\"{a}, Finland), and to 
study the impact of CCTV cameras on users' route planning when privacy or safety is a crucial factor. 
Our present work is different yet complementary to these studies. 
On the one hand, Sintonen et al.~\cite{sintonen2021osrm} and Lahtinen et al. ~\cite{lahtinen2021towards} provide pro-active planning PET tools, 
while our present work provides re-active (real-time or historical) tools, with further possible applications in digital forensics and investigative fields. 
Nevertheless, our present work is complementary as it optimally extends and complements~\cite{lahtinen2021towards,sintonen2021osrm} 
in building complete PET toolsets (i.e., a combination of pro-active and re-active) applicable to CCTV cameras, anti-surveillance, and public safety. 

Using GPS data to measure human exposure in different cases is nothing new to the general research field. 
Rinzivillo et al.~\cite{rinzivillo2013have} proposed a diary creation helper tool based on GPS data recorded during the day. 
Their tool allowed users to pinpoint their activities with temporal and spatial accuracy provided by accurate GPS data. 
Dias and Tchepel~\cite{dias2014modelling} used GPS data collected from test subjects' mobile phones to measure the users' exposure to air pollution. 
Their study was conducted in the Leiria area in Portugal, and their pollution data were estimated via Transport Emission Model for Line Sources (TREM) model and meteorological data. 
Dias and Tchepel claim that due to pollution concentration variation within ``microenvironments'', their exposure model will yield a meaningfully better understanding of individual's pollution exposure in urban areas in contrast to traditional background pollution measurements. 
Correspondingly, Tchepel et al.~\cite{tchepel2014modeling} measured human exposure to benzene in the Leiria area in Portugal. 
Several other studies (such as~\cite{beekhuizen2013performance,breen2014gps,ma2020assessing}) have also measured exposure to air pollutants using GPS data.
Signal et al.~\cite{signal2017kids} utilized GPS and wearable cameras in a study in Wellington, New Zealand, to study children's exposure to food and beverage marketing. They recruited 168 children to wear the devices and record data over four days. Their method provided insight into food marketing exposure from a child's perspective while eliminating any observing researcher's biases.

Global positioning system data are also valuable for creating large datasets of human mobility data. 
These datasets can be used in conjunction with machine learning and artificial intelligence technologies, for example, to predict crowd flows. 
Luca et al.~\cite{luca2020deep} surveyed on that subject. 
However, they concluded that at the time of publishing in December 2020, state-of-the-art models for predicting human mobility suffer from several limitations, for example, data privacy concerns and the geographical constraints for the trained models. 


Global positioning system devices and data sending units are also used in tracking wildlife. 
Hinton et al.~\cite{hinton2019gps} measured Cesium-137 exposure on wildlife in the Chernobyl exclusion zone in Ukraine from November 2014 to May 2015. 
They attached a GPS monitoring unit and a dosimeter to eight free-ranging wolves in the area for data gathering. 
The gathered dosage data was used to analyze the soil Cesium-137 levels in relation to the temporal and spatial data collected from the GPS units.


\section{Design and Implementation}
\label{sec:impl}

\begin{figure}
\includegraphics[width=\textwidth]{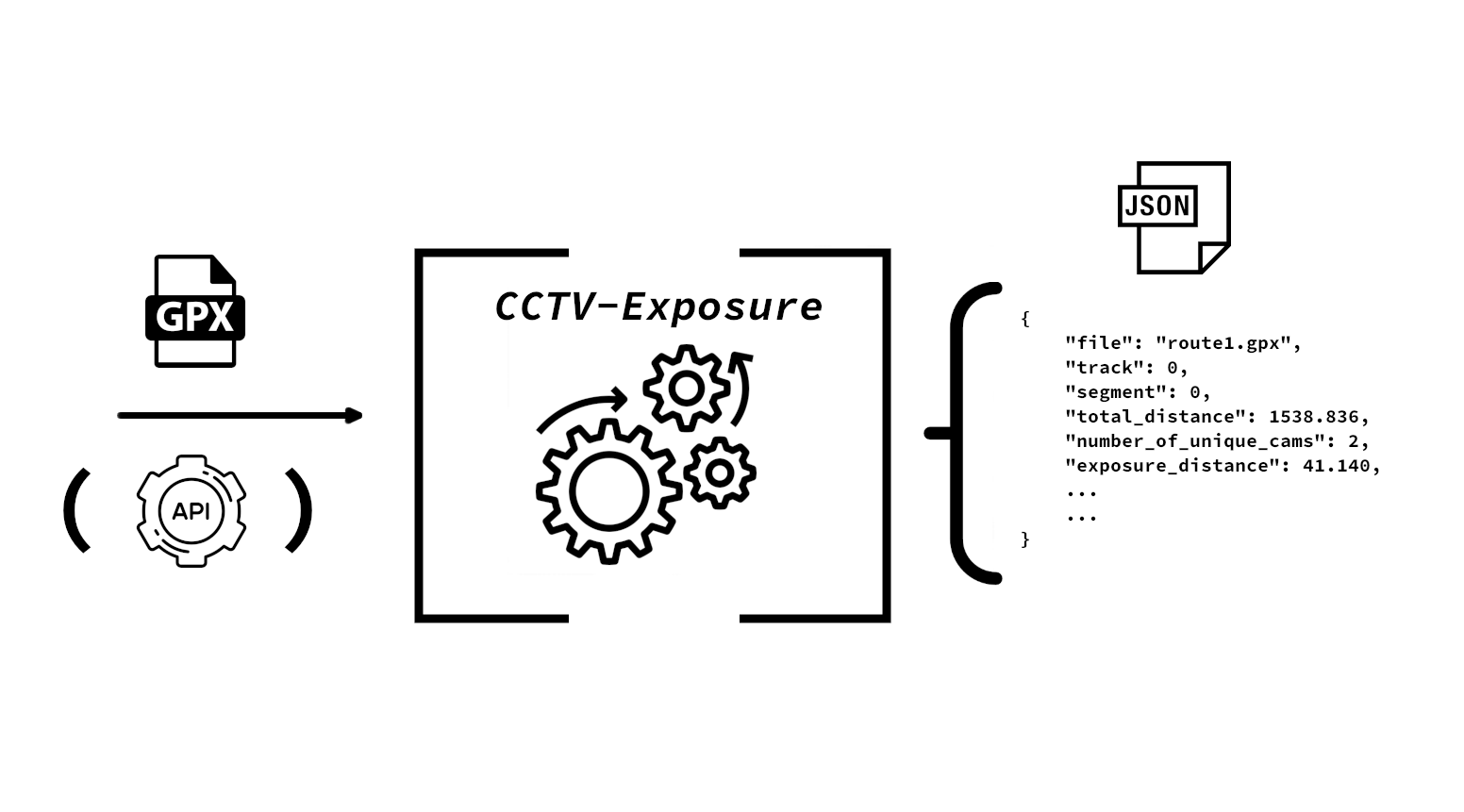}
\caption{\cctvexposure{} system overview.}
\label{fig:exp_system}
\end{figure}

Our \cctvexposure{} system (illustrated in Figure~\ref{fig:exp_system}) is written in Python3 with minimal requirements, as only 
GPXpy~\footnote{ \url{https://pypi.org/project/GPXpy/}, \url{https://github.com/tkrajina/GPXpy} } and 
NumPy~\footnote{ \url{https://numpy.org/} } are used to reduce any code and dependency overheads. 
GPXpy is a global position system exchange (GPX) file parser by Tomo Krajina. 
It is a simple yet effective way to efficiently parse extensible markup language (XML)-based GPX files. 
For trigonometric operations with geo-coordinates, we opted to use NumPy for the calculations instead of utilizing previously written projects or the standard Python math library. 

We also implemented the module in Rust (v. 1.60). The Rust implementation is similar and equivalent to our Python3 counterpart; however, it does not allow the use of non-timestamped GPX files due to parser limitations.

At present, and for this paper's evaluation, our system accepts only GPX files as input. However, an application programming interface (API) input is an option that we leave to be implemented in future work. 
In the case of multiple files or large files with several tracks and routes, the module can be adapted easily to apply parallel multi-processing paradigms (e.g., utilize various processes for each CPU core) to take advantage of available processing power and save time-to-results. 
Another required argument for the module is the camera database file. For our testing, we used the camera coverage radius and the field-of-view specified in the camera database file; however, these values can be overridden with input arguments by the user. It enables fast prototyping, experimentation, and re-evaluation.

We decided to use Euclidean distance to perform faster computations instead of the more accurate Haversine distance. 
However, the module allows option specification to easily switch between Euclidian and Haversine distances and add alternative distance measurement implementations. 
The core module performs all calculations in meters (for distance) and seconds (for time). 
For this reason, we ignore the curvature of the earth to quicken the calculations, as well as simplify the model. 
Our synthetic tests show that for GPX tracks of several kilometers, the cumulative error is negligible when assuming realistic (e.g., hundreds of meters to several kilometers) human geo-location tracks within CCTV-fitted public spaces. 
For example, we calculated an inaccuracy of one meter within a range of over 700 meters between the different distance calculations in distance calculations. As we deal with calculations of 50 meters and less, the extra computational effort of the more accurate calculations is not warranted, at least at this early-research stage. 
At the same time, the display and output module can be adapted for specific locales and units of measurements to accommodate a global audience; however, this does not affect the core computations.

The general operation of our module is presented in Algorithm~\ref{alg:1}. 
The gist of the module is to loop over all tracks and segments read from the GPX input file. 
It is important to note that the input GPX file can (and should) be wholly anonymized and scrubbed of any Personally-Identifiable Information (PII), as \cctvexposure{} aims to enhance and preserve privacy as one of its core principles. 
There are many practical methods that an individual non-tech-savvy user of \cctvexposure{} could create an anonymized and scrubbed GPX file, even if it comes from their personalized history track and GPS-enabled device. 
A GPX file can have multiple tracks, which can have multiple segments, and each segment is specified in the GPX file via a set of points (i.e., exact GPS locations, with optional timestamp). 
We refer to these points throughout the rest of this paper as \emph{GPX points}. 
The core module loops over each GPX point and identifies ``in-range cameras'' for each iterated point. 
An ``in-range camera'' for a GPX point means a CCTV camera (from the available and loaded database of geo-location mapped CCTV cameras) whose field-of-view (whether directed or 360) covers or intersects with the GPX point. 
Afterward, the module:
\begin{itemize}
    \item loops all GPX points which we concluded above to be ``within the visual reach'' of their ``in-range cameras.''
    \item splits the distance between the point and their adjacent point for more granular inspection and calculation (see Section~\ref{sec:points})
\end{itemize}
For this proof-of-concept module, we loaded our database of geo-location mapped CCTV cameras from a JSON file and then looped over all of them for each point. 
However, for larger dataset operations (e.g., EU-wide CCTV camera database) this would prove highly sub-optimal, therefore, for any operation beyond current evaluation (i.e., Jyv\"{a}skyl\"{a}, Finland), 
a production-grade implementation would benefit from a high-performance database/APIs and a smart search algorithm to loop over all the necessary but a minimal number of cameras for each point. 
We leave this performance optimization challenge for future work. 

\begin{algorithm}
\begin{algorithmic}[1]
\caption{Generalization of \cctvexposure{} data processing.}
\label{alg:1}
\Procedure{main}{$GPX$}:
    \State load\_GPX
    \State load\_cameras
    \ForAll {tracks in GPX}
        \ForAll {segment in tracks}
            \State initialize
            \State calculate\_exposure
            \State create\_output
        \EndFor
    \EndFor
\EndProcedure
\newline

\Procedure{initialize}{$segment$}
    \State get\_total\_distance
    \If{time in segment}
        \State time\_present
        \State get\_total\_time
    \EndIf
    \ForAll {points in segment}
        \If{speed is present}
            \State speed\_present
        \EndIf
        \State calculate\_distance\_to\_cameras
            \If{in\_range\_of\_cameras}
                \State save\_cameras\_info
            \EndIf
    \EndFor
\EndProcedure
\newline

\Procedure{calculate\_exposure}{$points\_with\_cameras$}    \Comment{repeat}
    \ForAll {points in points\_with\_cameras}   
        \ForAll {camera in points}
            \State calculate\_distance  \Comment{previous or next point}
            \State calculate\_bearing \Comment{to previous or next point}
            \If{same camera in adjacent point}
                \State $interpolated\_points$ = $distance\_to\_previous$ / $resolution$
            \Else
                \State calculate\_interpolated\_points \Comment{detection beyond points}
            \EndIf
            \If {speed\_present or time\_present}
                \State calculate\_exposure\_time
            \EndIf
            \State save\_highest\_coverage
            \State save\_individual\_cameradata
            
        \EndFor
    \EndFor
\EndProcedure
\newline

\Procedure{create\_output}{results}
    \State calculate\_statistics    \Comment{averages, percentages}
    \State create\_json\_output    
\EndProcedure
\newline

\end{algorithmic}
\end{algorithm}

\subsection{Interpolated points - points between GPX points}
\label{sec:points}

\begin{figure}[htb]
  \centering
  \subfloat[\cctvexposure{} interpolated point system: blue markers are GPX points and red markers are interpolated points between them. \label{fig:interpolated}]{\includegraphics[width=.475\textwidth]{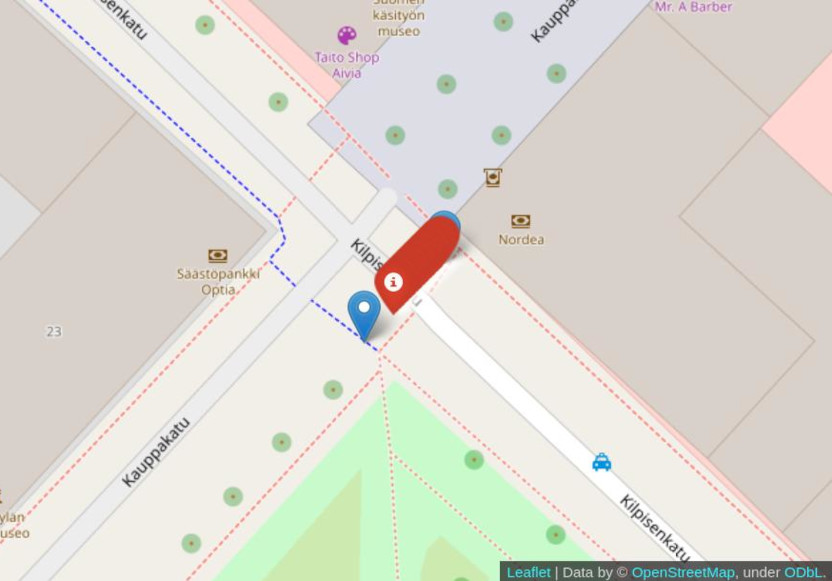}}\quad
    \subfloat[\cctvexposure{} interpolated points measured: interpolated points in yellow, camera in red, GPX points in blue.\label{fig:interpolated2}]{\includegraphics[width=.475\textwidth]{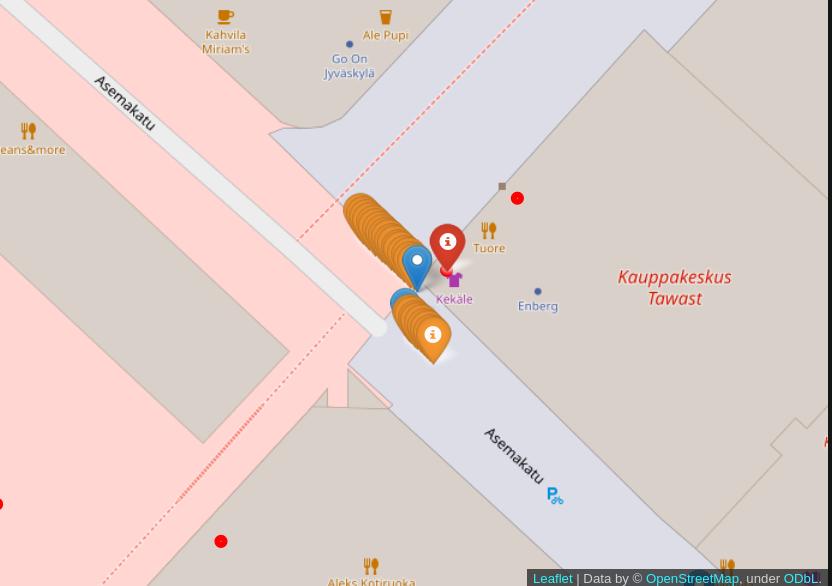}}\quad
  \subfloat[\cctvexposure{} granular calculation: double-sided arrows mean that the granular calculation is not required, while single-sided arrows mean granular calculation is required. \label{fig:backtracking}]{\includegraphics[width=.75\textwidth]{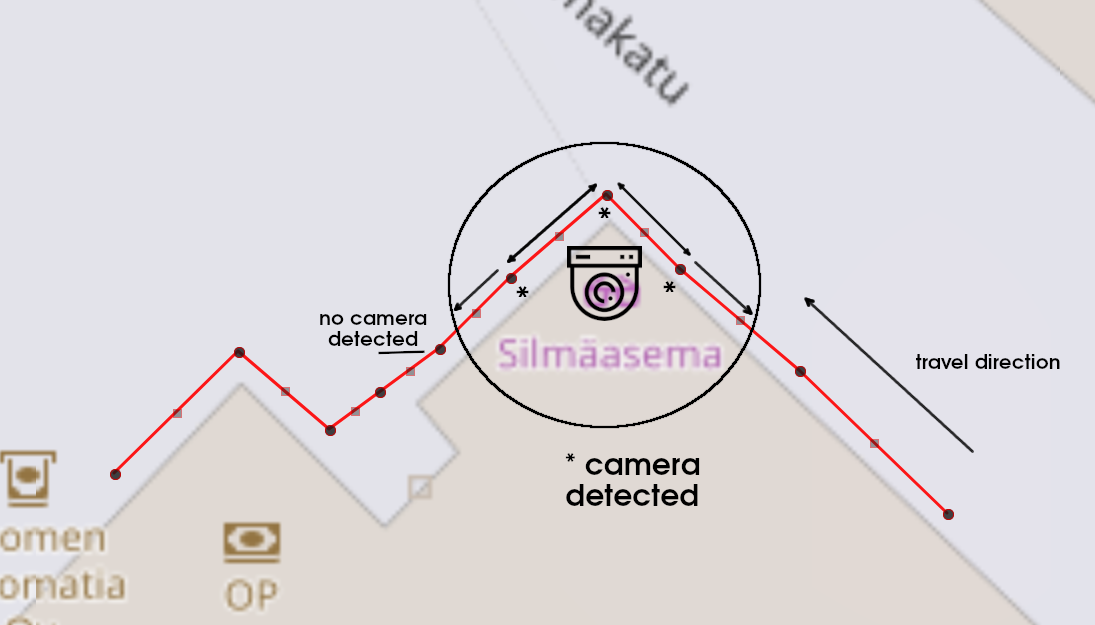}}\\

  \caption{Interpolated point depiction.}
  \label{fig:sub1}
\end{figure}

To increase the accuracy of our calculations, for each GPX point that a camera is in range of, we split the distance to \emph{interpolated points}. 
\emph{Interpolated points} are not present in GPX data and are a result of our internal calculations to increase both the granularity of analysis and accuracy of the exposure estimate. 
A variable splits the distance we call \emph{resolution}. 
The default resolution for our experiments was 0.5 meters; however, the resolution can configure into the system for lower or higher granularity and accuracy purposes. 
This value is a critical parameter to adjust due to the inherent uncertainty of the GPS data and the variability of data accuracy (GPS drift). 
The interpolated points are illustrated in Figure~\ref{fig:interpolated}. 
We skip this calculation if the distance between GPX points is less than the resolution or if a particular camera is present with the previous GPX point and the current one during the loop.
If a particular camera is present with the previous GPX point and the current one during the loop, we skip the interpolated point calculation. 
The skip can be made because a route between GPX points is a straight line; therefore, we can be sure that the camera covers the whole distance. 
Otherwise, we go through the GPX points with cameras and calculate how far back from the present GPX point the field-of-view of the camera in question reaches (i.e., how many interpolated points). 
From the answer, we can accurately (within resolution) calculate the distance and time spent in the field-of-view of the individual camera. 
We save each camera coverage and the maximum camera coverage for that point (i.e., the highest number of interpolated points for the GPX point). 
This highest number adds to the total exposure to any camera across the whole route. 
However, we think that individual camera data can also be valuable.

After the backward iteration, we loop the GPX points again but focus on going forward from the GPX points, where each camera stops in range. This way, we cover the whole union of camera field-of-view and the route in the process. Figure~\ref{fig:backtracking} stages the idea behind the back-and-forth coverage system. Also in Figure~\ref{fig:interpolated2}, calculated interpolated points are demonstrated. In the example, both blue GPX points are in the red camera's range; therefore, there's no need for granular calculation between the points.

\subsection{Module output}

Our module results in JavaScript Object Notation (JSON) formatted output with the following items:

\begin{itemize}
    \item Identity: identity information for the processed file, track, and segment
    \item Distance: total distance traveled in the segment, total exposure distance, average and mean distance to cameras (in GPX points)
    \item Time (if applicable): average speed, total segment time, exposure time
    \item Percentages: exposure per total distance, exposure time per total time (if applicable)
    \item Per camera data: time, distance, and all camera data available (location, field-of-view, etc.)
    \item Number of unique cameras
\end{itemize}

Time-related output is only available if the points have speed information or timestamps in all GPX points per segment. 
A JSON-type output is preferable as it is flexible and can be easily saved as a file, saved to the database as a whole or as individual columns, or returned via API for further processing or dashboard display. 

In Listing~\ref{alg:2}, we present a sample JSON output.
\begin{lstlisting}[caption=Sample JSON output generated by \cctvexposure{} module. , firstnumber=1,label={alg:2}]
{
    "file": "route2.gpx",
    "track": 0,
    "segment": 0,
    "total_distance": 1538.83,
    "number_of_unique_cams": 2,
    "exposure_distance": 41.14,
    "dist_percentage": 2.67,
    "camera_distance_avg": 1069.96,
    "camera_distance_median": 871.58,
    "avg_speed": 4.99,
    "time_percentage": 5.14,
    "exposure_time": 57.08,
    "cameras": {
        "133": {
            "latitude": "62.2415517",
            "longitude": "25.7452791",
            "camera_type": "round",
            "radius": "10.0",
            "angle_of_view": "360",
            "camera model": "Zmodo ZP-IBH23-S",
            "url": "",
            "camera_in_streetview": "no",
            "time_in_camera_fov": 32.92,
            "distance_in_camera_fov": 23.64
        },
        "199": {
            "latitude": "62.2438628",
            "longitude": "25.7500291",
            "camera_type": "directed",
            "radius": "10.0",
            "angle_of_view": "360",
            "camera_model": "Zmodo ZP-IBH23-S",
            "url": "",
            "camera_in_streetview": "yes",
            "time_in_camera_fov": 24.15,
            "distance_in_camera_fov": 17.5
        }
    }
}
\end{lstlisting}


\section{Evaluation and Results}
\label{sec:results}

The Jyv\"{a}skyl\"{a} city area in Central Finland was chosen as the experiment location for this study. 
As of writing, it is the seventh-largest city in Finland by population. 
The immediate city center area is relatively compact and rather CCTV congested. 
The camera mapping was conducted in the summer of 2020~\cite{lahtinen2021towards,sintonen2021osrm}, and the routes for this study were captured in early 2022. 
Even if the camera mapping might have changed in-between, the purpose of this work and evaluation is the development, demonstration, and evaluation of the system and its implementation (including performance and computation). 
Our system can be re-run with ease once updated camera mappings become available, thus offering a historical perspective on the change of potential CCTV exposure over time on the same track (e.g., allows monitoring and evaluation of CCTV privacy invasion evolution). 
For this evaluation section, we have used peer-review accepted approaches and methodologies from adjacent related works~\cite{dias2014modelling,tchepel2014modeling}.

\subsection{Evaluation Methodology}

\begin{figure}[htb]
  \centering
  \subfloat[Example of a synthetic route traveling through the narrow side of the city.]{\includegraphics[width=.4\textwidth]{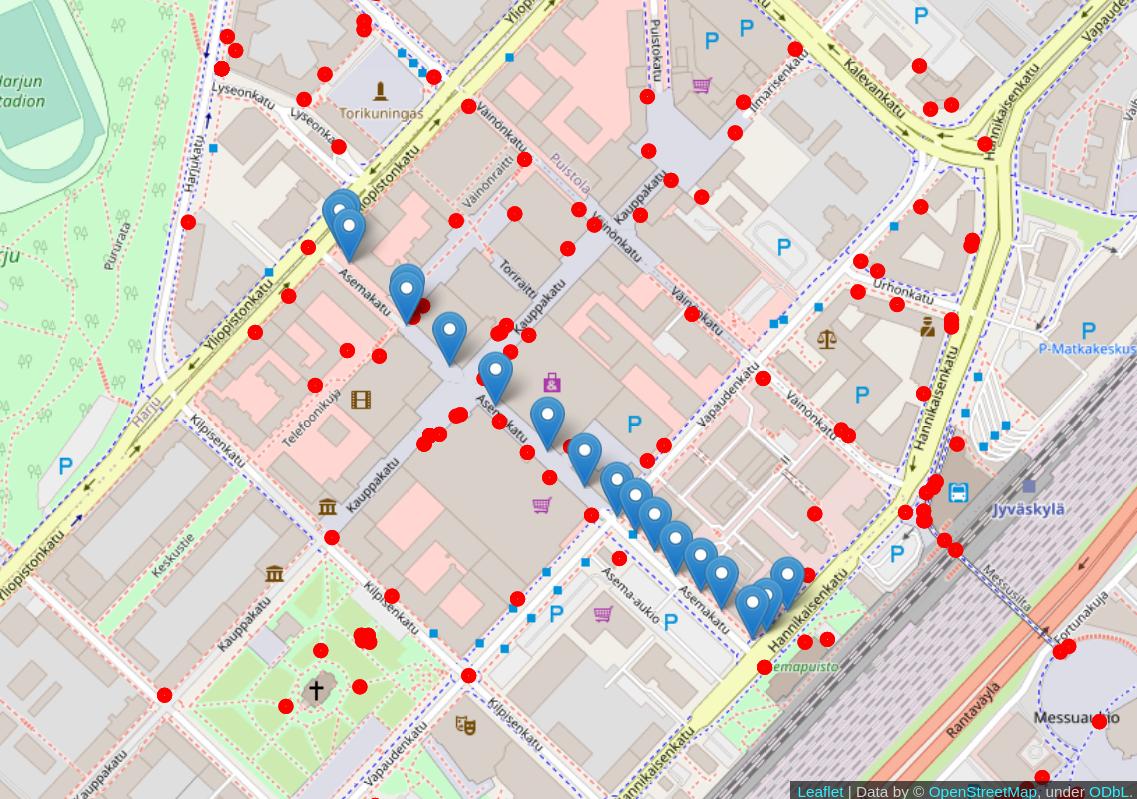}}\quad
  \subfloat[Example two of a synthetic route traveling alongside the city center]{\includegraphics[width=.4\textwidth]{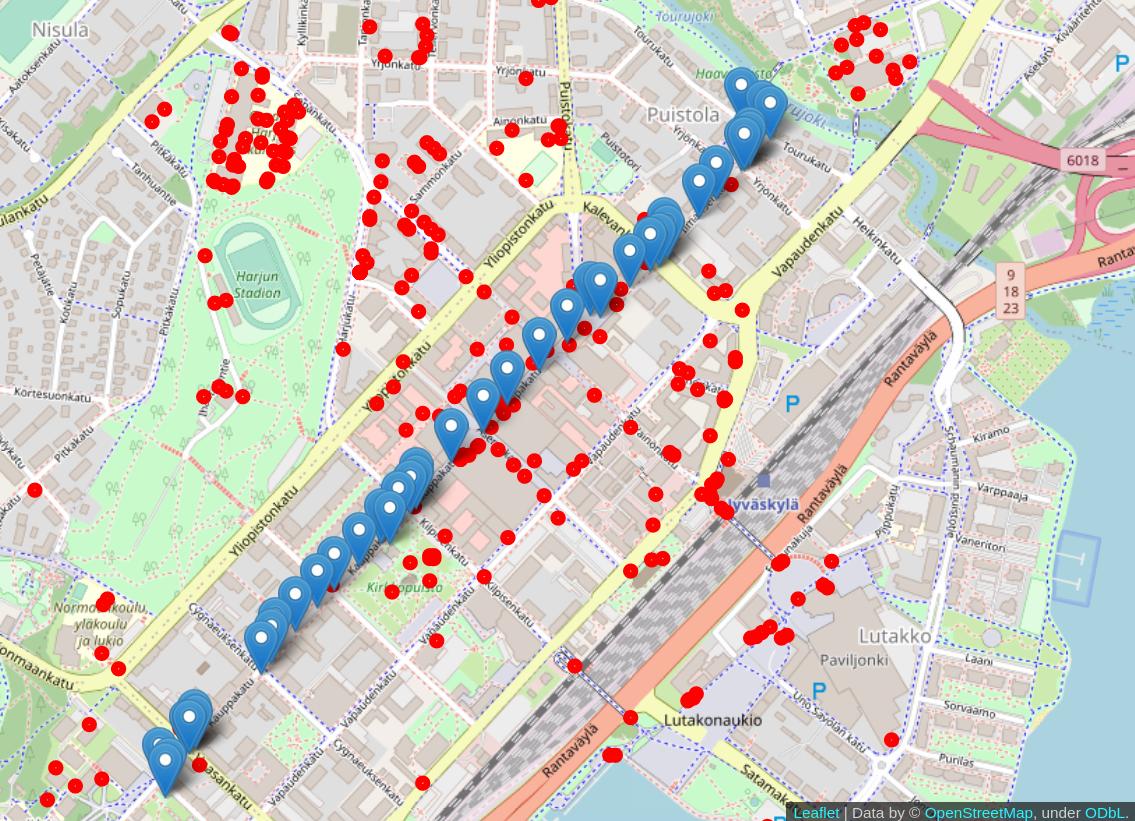}}\\
  \subfloat[Example of a recorded route in the city center.]{\includegraphics[width=.4\textwidth]{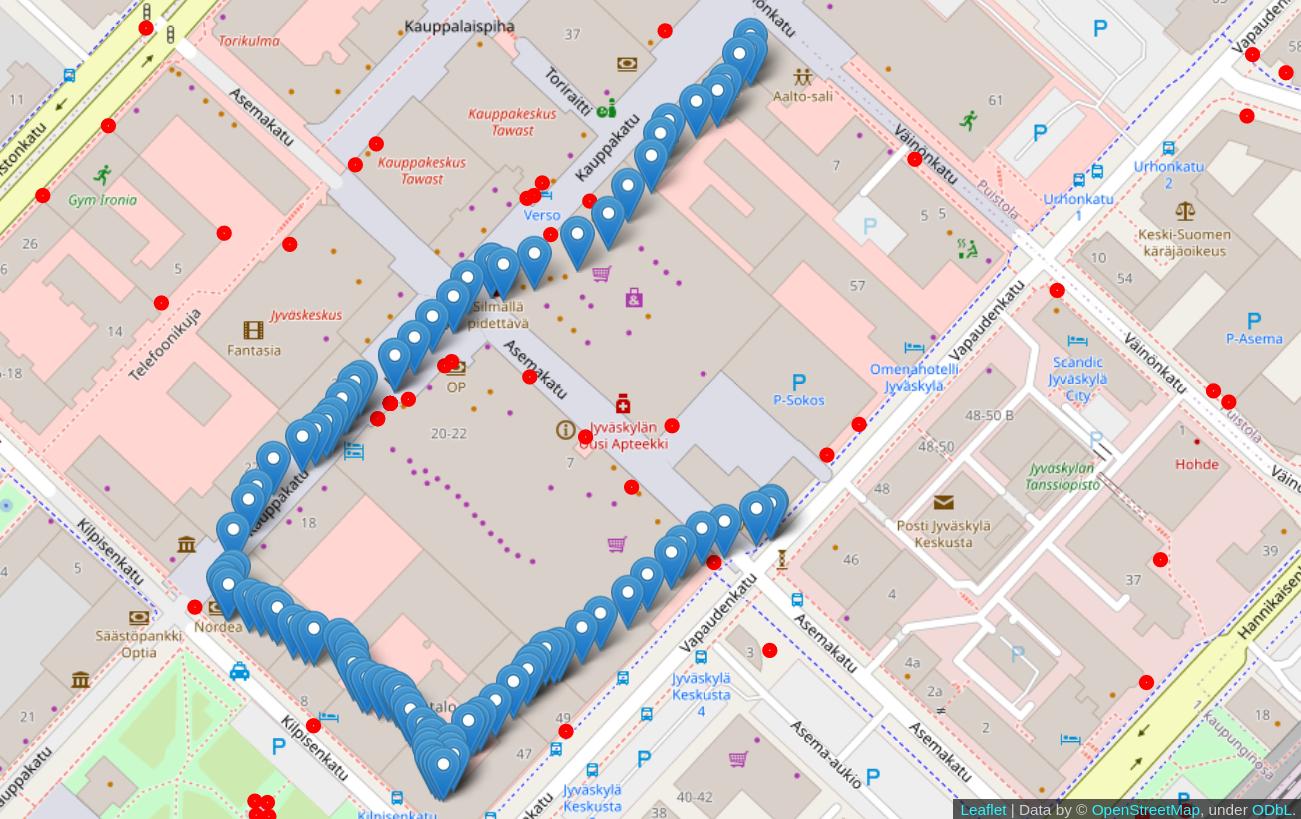}}\quad
  \subfloat[Example two of a recorded route in the city center.]{\includegraphics[width=.4\textwidth]{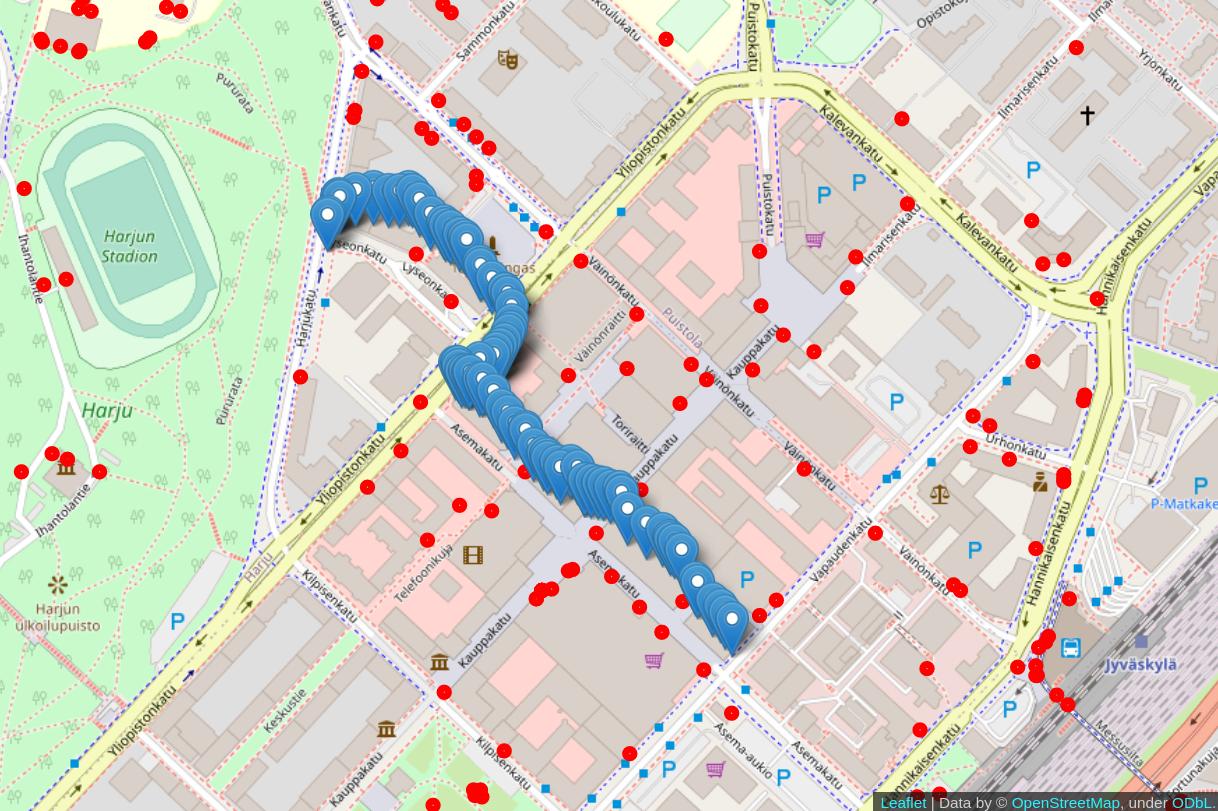}}
  \caption{Example routes used in evaluation of \cctvexposure{}.}
  \label{fig:routes}
\end{figure}

We tested four tool-generated 'synthetic' GPX files, and four Garmin recorded 'real-world scenario' files as routes for evaluation of \cctvexposure{}. 
Map examples of the routes are depicted in Figure~\ref{fig:routes}, where the routes are marked with blue markers and cameras with red dots. 
The 'synthetic' GPX files were created using GPSVisualizer~\footnote{ \url{https://www.gpsvisualizer.com} }, and timestamps were added to them using GoToes GPX editing tool~\footnote{ \url{https://gotoes.org} }. 
The timestamps followed a preset average speed. 
Our Garmin EDGE 810 recorded the files in Garmin's FIT format, which were subsequently converted into GPX using a converter from AllTrails~\footnote{ \url{https://www.alltrails.com/converter} }. 
The real-world scenario files provide a bit more exciting data as the timestamps are more varied due to changes in the recorded speed of the person; therefore, the exposure time and distance will yield differing results. 
However, the Garmin device used in this test had some accuracy issues during recording. 
This is unsurprising as GPS in urban areas can have inaccuracies due to high buildings and canopies~\cite{merry2019smartphone}, also known as the ``urban canyon'' effect on GPS. 

The real-world recordings were random walks around the city center while trying to include different areas. 
The recordings included changes in pace and some random stoppages; therefore, we expect minor differences in our results' exposure time and distance percentages. 
Our device did not include speed data in our recordings; thus, we need to rely on the average speed system between GPX points while calculating the interpolated point exposure metrics. 
The lack of speed data will inherently lead to added inaccuracies in the final exposure results. However, as we cannot create better data within the GPS device itself, minor inaccuracies must be tolerated. 
Nevertheless, these inaccuracies have less impact than ground contour deviations and GPS drift. 

\subsection{Evaluation Environment}

\begin{figure}[htb]
\centering
\includegraphics[width=0.95\textwidth]{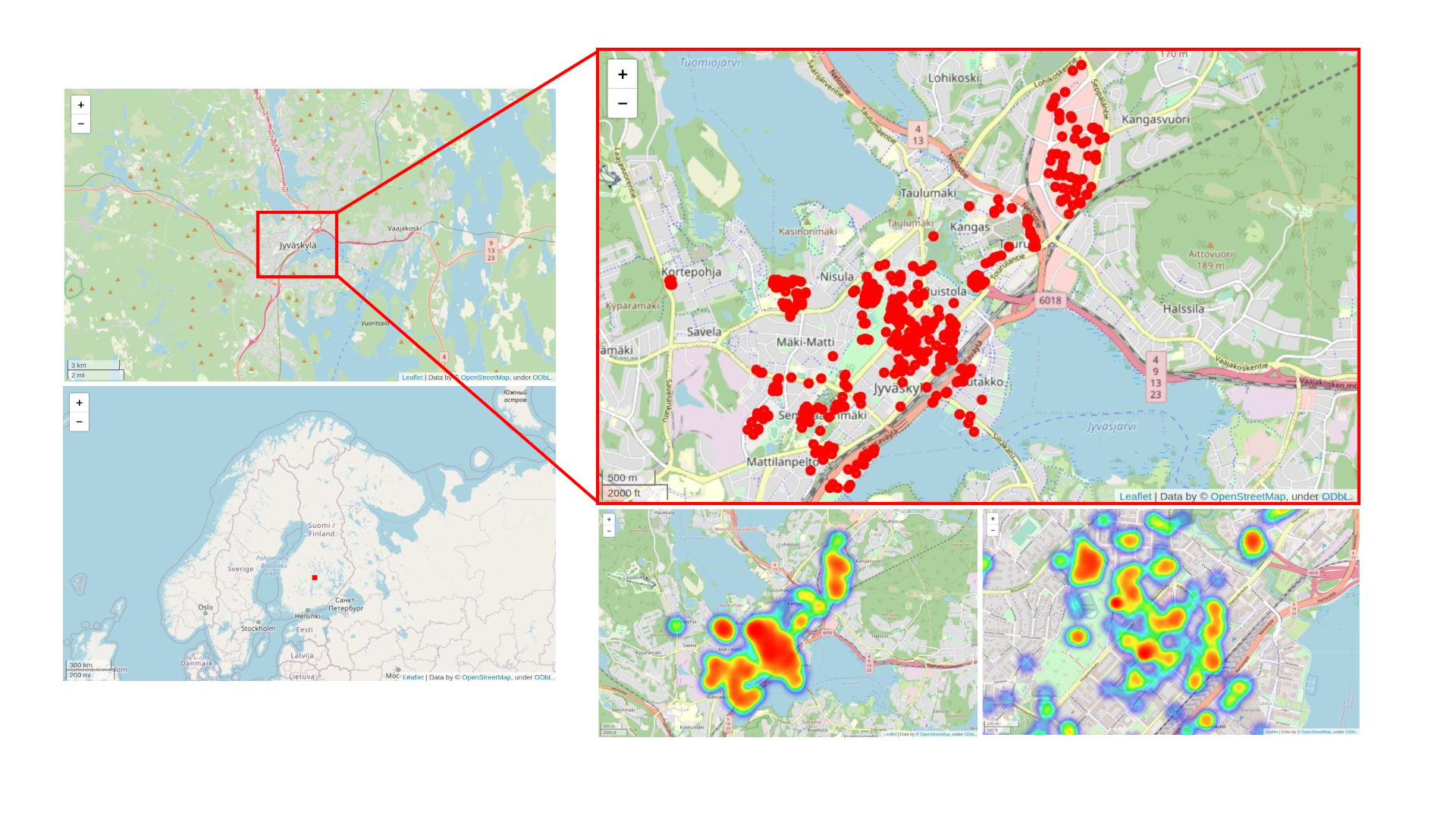}
\caption{CCTV cameras dataset mapped in the city of Jyv\"{a}skyl\"{a}~\cite{lahtinen2021towards}, and used for evaluation of \cctvexposure{}.}
\label{fig:jkl_cams_heatmap}
\end{figure}

In our current test setup, we were limited by our camera dataset containing only 450 cameras~\cite{lahtinen2021towards} around the city center of Jyv\"{a}skyl\"{a} (Finland). 
However, there is an active work-in-progress to expand this dataset rapidly in various parts of the world. 
For example, in an ongoing collaboration with a Dutch research group that works closely in the same field as our group, we are contributing to the creation of geo-location CCTV camera mapping for the city of Amsterdam (Netherlands), where the final goals are very similar to the plans of \cctvexposure{}. 
Nevertheless, any city will work in our \cctvexposure{} system as long as the cameras are mapped through our project, or the geo-location mapped CCTV cameras database is provided in a machine-readable format featuring minimally required data fields so that \cctvexposure{} can ``convert and import'' the data. 
Figure~\ref{fig:jkl_cams_heatmap} shows the mapped cameras that were used in the experiments. 

\subsubsection{Certain Assumptions and Limitations}

For our testing, for all the cameras in our database, we set by default ten (10) meters for the camera ``privacy invasion'' radius, 
i.e., the radius on which we assume any CCTV camera in the database can successfully record hard- and soft-biometrics of an individual with subsequent potential recognition or identification. 
This radius setting is highly conservative and emulates the ``worst-case scenario'' (i.e., limited visibility range from a CCTV camera perspective). 
In this range, most of the deployed and in-use cameras still can provide recordings where an accurate person identification is likely~\cite{wheeler2010face}. 
However, modern CCTV cameras can feature a ``privacy invasion'' radius from 20 to hundreds of meters, depending on the optical zoom and sensor specifications. 
At the same time, it means that the exposure to CCTV results presented below is a ``best-case scenario'' (i.e., lower-boundary exposure levels) for the individual user seeking privacy protection. 
In reality, we expect individual users to have their camera invaded and CCTV exposure at a much higher rate, assuming CCTV cameras have superior optical zoom and 360-degree lenses. 

Moreover, our CCTV camera dataset (and any other public dataset we have seen) has limitations as these datasets do not have 100\% accurate characteristics of each camera in the dataset. 
One core reason for this is that we can detect the presence of the camera (e.g., using CCTVCV~\cite{turtiainen2020towards} or crowdsourcing); however, we (and any similar third-party project) will not know certain information about each camera, such as:
\begin{enumerate}
    \item exact camera model -- this would also be challenging to perform visually by humans (due to low resolution and lack of markings) and via computer vision (as this would require the equivalent of ``face recognition'' accuracy and system, but for CCTV cameras).
    \item exact owner/operator of the camera(s) -- these contacts are generally missing but could (or perhaps \textbf{should}, as required by GDPR?) provide much more meta-information about the camera; we have a work-in-progress towards achieving this meta-information collection via crowdsourcing; however, we leave this challenge as future work.
\end{enumerate}

\subsection{Exposure results}

In Table~\ref{tabl:results}, we disclose the exposure metrics for the test routes.
Our synthetic routes' average CCTV exposure distance and time were 3.7\% of the total routes. 
Our real-world routes (captured by the authors with Garmin devices in the city center of Jyv\"{a}skyl\"{a}) indicated an average CCTV exposure of 12.5\% relative to segments' distance, and 15.1\% close to segments' time, respectively.

\begin{table}[]
    \centering
    \begin{tabular}{|l|l|l|l|l|}
    \toprule
    \textbf{Route} & \textbf{Distance -- m} & \specialcell{\textbf{Unique cams -- num}} & \specialcell{\textbf{Exposure dist. -- \% }} & \specialcell{\textbf{Exposure time -- \% }} \\ 
    \hline
syn1.gpx & 1538 & 2 & 2.6\% (41m / 1538m) & 2.6\%   (0:00:29 / 0:18:29) \\
syn2.gpx &  497 & 1 & 3.7\% (18m / 497m) & 3.7\%  (0:00:20 / 0:09:15) \\
syn3.gpx &  571 & 1 & 3.5\% (20m / 571m) & 3.5\%  (0:00:14 / 0:06:50) \\
syn4.gpx &  897 & 3 & 4.8\% (43m / 897m) & 4.8\%  (0:00:32 / 0:11:08) \\
    \hline
real1.gpx & 633 & 13 & 25.4\% (161m / 633m) & 36.0\%  (0:02:55 / 0:08:05) \\
real2.gpx & 614 & 14 & 17.4\% (107m / 614m) & 17.4\%  (0:01:16 / 0:07:15) \\ 
real3.gpx & 597 & 9 & 4.7\% (28m / 597m) & 4.5\%  (0:00:18 / 0:06:41) \\
real4.gpx & 775 & 1 & 2.3\% (18m / 775m) & 2.6\%  (0:00:15 / 0:09:17) \\
    \bottomrule
    \end{tabular}
    \caption{Exposure results for the routes used during preliminary evaluation.}
    \label{tabl:results}
\end{table}

Our synthetic routes had a more sparse point distribution, thus resulting in lesser exposure metrics. 
Our real-world recordings, however, produced more accurate data. 
Initially, our module looked for in-range cameras within the GPX points; we relied on these experiments because the capturing devices would frequently create data points. 
For example, with Garmin GPS recording devices, the user can choose whether the device records a point every second or when it detects a change in course~\cite{Garmin_help}. 
As this option is software-bound, the user has a certain degree of choice in this regard.

From a code performance standpoint, our module was run as a single-core task on an Intel i7-6700K running at 4.4GHz using perf. The time was taken as an entire run with full GPX and camera file loading times included with only one segment per run. Therefore, using the module with multiple files or numerous segments will yield better results per segment as these file loadings are only done once. On average, \cctvexposure{} took for Python3 0.169 seconds and for Rust 0.004 seconds to run each of our test routes by iterating over all of our 450 cameras on each GPX point (where the average distance across all tested routes is 766 meters). By excluding the file loading times from the results, our Python version clocked in at 0.06 seconds on average per segment, thus raising the question if our own GPX file parser would be beneficial from a performance standpoint.

More intelligent decisions on which cameras to iterate over will be needed in future work when our camera dataset increases. 
At the same time, the realistic routes for an average individual would not span city boundaries; therefore, the dataset of memory-loaded and iterated CCTV cameras can permanently be restricted to certain geo-fence limits where the GPX track was captured. 
We also tested the performance with perf on a Raspberry Pi 4 device, on which \cctvexposure{} took on average for Python3 1.07 seconds and for Rust 0.0156 seconds to run precisely the same routes over the same dataset of 450 cameras. 
All of the routes were run ten times to produce the averages.
These measurements increase when more cameras are on route as more interpolated point calculations are required. 
However, these results indicate that \cctvexposure{} Python3 could operate quite fast even on older or on mobile/IoT hardware despite being written in Python, which is generally considered slow as being an interpreted programming language. For multiple files and files with a large number of segments, using a ``just-in-time'' compiler for Python, such as Numba~\cite{numba}, could be beneficial. For our test cases, however, using it slowed the execution. 

Our Rust version is fast from the get-go, and it is undoubtedly the preferred choice to use as a standalone tool; however, the Python version is quite handy with project integrations. 
On the other hand, our Rust version would be ideal for longer GPX segments such as those recorded from car drives.

Based on these results, and especially on the real-world recorded routes, it is pretty safe to say that avoiding CCTV cameras around the city center of Jyv\"{a}skyl\"{a} (Finland) proves to be a challenge. 
These observations are in line with the conclusions from Lahtinen et al.~\cite{lahtinen2021towards}, where the authors implemented and studied CCTV-aware route-planning for ``preventive privacy analysis'', 
using an early prototype of OSRM-CCTV~\cite{sintonen2021osrm} as a CCTV-aware navigation and route planning solution.

It is important to note, however, that due to the systematic lack of previous works, ground truth datasets, and baseline recommended exposure levels related to ``CCTV privacy invasion'', 
the evaluation numbers should be interpreted with care because they represent a best-effort estimate of the privacy exposure to the CCTV cameras based on the limited CCTV camera datasets and the error-prone GPS tracks. 
This systematic lack of fundamental support in this field also means more work is required at multiple levels (e.g., research, baseline standardization, policy, technology adoption, and promotion).


\section{Conclusion}
\label{sec:concl}

In this paper, we presented \cctvexposure{} -- an open-source system for measuring users' privacy exposure to mapped CCTV cameras based on geo-location, GPX, and historic tracks. 
We implemented \cctvexposure{} system prototype as a server-side module that is easy to expose via API for easy integration into more comprehensive and more user-friendly solutions. 
We evaluated the \cctvexposure{} on multiple GPS tracks in Jyv\"{a}skyl\"{a}, where we also had a comprehensive CCTV camera mapping database of 450 cameras. 
Our evaluations demonstrate the effectiveness, performance, and practicality of \cctvexposure{} when tasked with measuring CCTV exposure of users based on their real-time or historical geo-location and GPS tracks. 

As this is some early yet promising implementation and evaluation, certain limitations and challenges have been identified. They present a fertile ground for further research that we leave as future work. 
First, performance optimization and accuracy validation of \cctvexposure{} will benefit from collecting more extensive and more complete CCTV databases while being validated on larger and more diverse datasets. 
Second, \cctvexposure{} will benefit from being validated with larger and more diverse user-base and stakeholders' scenarios, which could bring insights into both necessary performance optimizations as well as user-experience and usability issues preventing achieving usable privacy technologies. 
Third, \cctvexposure{}, as well as OSRM-CCTV, would both benefit from a holistic integration into an end-to-end CCTV-aware system (including powerful and flexible API backends and user-friendly mobile apps and web interfaces), 
aimed at enhancing users' digital privacy at all stages -- route pre-planning, real-time positioning, and historical tracking. 

For evaluation and further improvements, as well as to encourage researchers and practitioners to explore this digital privacy-related field, 
we release (upon peer-review acceptance) the relevant artifacts (e.g., code, data, documentation) as open-source: \url{https://github.com/Fuziih/cctv-exposure}

\bibliographystyle{splncs03}
\bibliography{cctve}

\end{document}